\documentclass[aps,prb,twocolumn,superscriptaddress,bibliography]{revtex4-2}
\usepackage{amsfonts}
\usepackage{mathrsfs}
\usepackage{amsmath}
\usepackage{color}
\usepackage{graphicx}
\usepackage{bm}
\usepackage{amssymb}
\usepackage{xspace}
\usepackage{epstopdf}
\usepackage{dcolumn}
\usepackage{longtable}
\usepackage{multirow}
\usepackage{float}
\usepackage{comment}

\usepackage[colorlinks=true, letterpaper=true, pdfstartview=FitV,  linkcolor=blue, citecolor=blue, urlcolor=blue]{hyperref}

\usepackage{lipsum} % used only for adding dummy text

\makeatother

\begin{document}

\title{Plasmons in a two-dimensional nonsymmorphic nodal-line semimetal}

\author{Jin Cao}
%\email{ygyao@bit.edu.cn}
\affiliation{Centre for Quantum Physics, Key Laboratory of Advanced Optoelectronic Quantum Architecture and Measurement (MOE), School of Physics, Beijing Institute of Technology, Beijing, 100081, China }
\affiliation{Beijing Key Lab of Nanophotonics \& Ultrafine Optoelectronic Systems, School of Physics, Beijing Institute of Technology, Beijing, 100081, China }
\affiliation{Research Laboratory for Quantum Materials, Singapore University of
	Technology and Design, Singapore 487372, Singapore}

\author{Hao-Ran Chang}
\email{hrchang@mail.ustc.edu.cn}
\affiliation{Department of Physics, Institute of Solid State Physics and Center for Computational Sciences, Sichuan Normal University, Chengdu, Sichuan 610066, China}

\author{Xiaolong Feng}
\affiliation{Research Laboratory for Quantum Materials, Singapore University of
	Technology and Design, Singapore 487372, Singapore}

\author{Yugui Yao}
%\email{ygyao@bit.edu.cn}
\affiliation{Centre for Quantum Physics, Key Laboratory of Advanced Optoelectronic Quantum Architecture and Measurement (MOE), School of Physics, Beijing Institute of Technology, Beijing, 100081, China }
\affiliation{Beijing Key Lab of Nanophotonics \& Ultrafine Optoelectronic Systems, School of Physics, Beijing Institute of Technology, Beijing, 100081, China }

\author{Shengyuan A. Yang }
%\email{shengyuan\_yang@sutd.edu.sg}
\affiliation{Research Laboratory for Quantum Materials, Singapore University of
	Technology and Design, Singapore 487372, Singapore}

%\date{\today}

\begin{abstract}
Recent experiments have established a type of nonsymmorphic symmetry protected nodal lines in the family of two-dimensional (2D) composition tunable materials NbSi$_x$Te$_2$. Here, we theoretically study the plasmonic properties of such nonsymmorphic nodal-line semimetals. We show that the nonsymmorphic character endows the plasmons with extremely strong anisotropy. There exist both intraband and interband plasmon branches. The intraband branch is gapless and has a $q^{1/2}$ dispersion. It is most dispersive and is independent of carrier density in direction normal to the nodal line, whereas along the nodal line, its dispersion is largely suppressed and its frequency scales linearly with carrier density. The interband branches are gapped and their long wavelength limits are connected with van Hove singularities of the band structure. We find that the single particle excitations are strongly suppressed in such systems, which decreases the Landau damping of plasmons. These characters are further verified by first-principles calculations on 2D NbSi$_x$Te$_2$. Interesting features in static screening of charged impurity are also discussed.
Our result reveals characteristic plasmons in a class of nonsymmorphic topological semimetals and offers guidance for its experimental detection and possible applications.
\end{abstract}
\maketitle

\section{Introduction}

In topological semimetals (TSMs), the Fermi surface is consisting of symmetry-protected band degeneracies, such that the low-energy electronic states can have distinct characters in their energy dispersion, pseudospin structure, and interband coherence~\citep{Chiu2016,Burkov2016,Yan2017,Ashvin2018}. This leads to many interesting physical properties, different from conventional metals or doped semiconductors. Clearly, the dimensions of the system as well as of the degeneracy manifold play important roles in the physics of TSMs.
For instance, in three dimensional (3D) materials, stable degeneracy manifolds may take the form of nodal points~\citep{WanXG2011,XuGang2011,YoungSM2012,WangZJ2012,Bradlyn2016,Weng2016,ZhuZM2016}, nodal lines~\citep{SYang2014,Weng2015,Mullen2015,Chen2015}, or nodal surfaces~\citep{Zhong2016,Liang2016,WuW2018}. In recent works, the possible protected band degeneracies were systematically classified for all magnetic space groups~\citep{EP_TYPE_II,EP_TYPE_III,EP_TYPE_IV,Tang2021,Tang2022}.

The TSM physics has also been actively explored in two-dimensional (2D) materials~\citep{FengXL2021}, which is another research focus in the past two decades~\citep{Pere2014,Zhanghua2017}. In fact,  graphene is a prominent example of 2D nodal-point TSMs~\citep{Geim2009}, and studies on graphene drove the whole field of topological materials~\citep{Kane2005}. Meanwhile, there has been a lot of interest in 2D nodal-line semimetals~\citep{FengBJ2017,gao2018epitaxial}. Many candidate materials were proposed~\cite{FengXL2021}. However, most of the proposals suffer one or more of the following drawbacks. (1) The nodal line is far away from the Fermi level. (2) Points on the nodal line have a large variation in energy. (3) There are other extraneous band coexisting at Fermi energy. (4) The material is hypothetical, or is not stable at ambient condition. Points (1-3) make it difficult to probe signatures of nodal-line states, and point (4) poses challenge for experimental studies as well as possible applications.

Recently, a series of theoretical and experimental works have established a good TSM state in the NbSi$_x$Te$_2$ family materials~\citep{Li2018_NbSiTe,Sato2018_NbSiTe,Yang2019_NbSiTe,Zhu2020_NbSiTe,Wang2021_NbSiTe,Zhang2022_NbSiTe}. These materials were first synthesized in the 1990s~\citep{li1992synthesis}. In the 3D bulk form, they are van der Waals layered materials, so 2D ultrathin layers can be readily obtained via mechanical exfoliation~\citep{Hu2015}. It was shown that the bulk material possesses hourglass nodal loops, whereas the 2D monolayer is an almost ideal nodal-line semimetal protected by a nonsymmorphic symmetry~\citep{Li2018_NbSiTe}. More interesting, this family belongs to so-called  composition-tunable materials, i.e., there exist a series of stoichiometric members with the formula Nb$_{2n+1}$Si$_n$Te$_{4n+2}$, or equivalently NbSi$_x$Te$_2$ with $x=n/(2n+1)\in[1/3,1/2]$~\citep{li1992synthesis,monconduit1993synthesis,evain1994modulated,van1994superspace}. While the specific band dispersion varies with $n$, all these members in 2D feature the nonsymmorphic nodal line~\citep{Zhu2020_NbSiTe}. It was found that the low-energy bands that make the nodal line are mainly from an array of NbTe$_2$ chains, which can be well captured by a 2D Su-Schrieffer-Heeger (SSH) like model~\citep{Zhang2022_NbSiTe}. Evidently, the NbSi$_x$Te$_2$ family materials offer a promising platform for studying the physics of 2D nodal-line TSMs.

Among various physical properties, plasmons, the collective modes associated with density oscillations of electron liquid, could directly manifest the distinct characters of low-energy electronic states. Indeed, plasmons of graphene and other TSMs have attracted great interest~\citep{Hwang2007,Hwang2008,DasSarma2009,Zhang2010,Zhang2013,Chang2014,Juergens2014,Zhou2015,Hofmann2015,Kharzeev2015,Sadhukhan2020}. However, plasmons associated with nodal-line states were mainly studied in 3D systems, and the line is often assumed to have the shape of a perfect ring in Brillouin zone (BZ)~\citep{Yan2016,Rhim2016,Wang2021a}. This is not the case for the nonsymmorphic nodal line in 2D NbSi$_x$Te$_2$, which traverses the BZ along a straight path and exhibits strong anisotropy~\citep{Zhu2020_NbSiTe,Zhang2022_NbSiTe}. Thus, it is desirable to find out how this nonsymmorphic nodal-line state impacts the plasmon properties.

In this work, we undertake this task and investigate the plasmons of nonsymmorphic nodal-line TSMs. First, by using the 2D SSH like model, we show that the plasmon spectrum contains two parts, denoted as intraband and interband plasmons. The intraband plasmon branch is gapless and has a $q^{1/2}$ dispersion. The scaling of plasmon frequency with carrier density $n$ (or Fermi energy) depends on the propagation direction, which crossovers from $\sim n^0$ normal to the nodal line to $\sim n$ along the line. Meanwhile, the two interband branches are gapped and are connected to the van Hove singularities of the band structure. All plasmon branches exhibit strong anisotropy and a characteristic angular dependence. The single particle excitations are strongly suppressed in such systems, such that the plasmons may enjoy less damping and longer lifetime.
By using first-principles calculations, we show that the key features obtained from the model study can indeed manifest in monolayer NbSi$_x$Te$_2$, which distinguish the system from conventional metals and doped semiconductors. In addition, we show that the screening charge density induced by a charged impurity also exhibits an interesting signature owing to the nonsymmorphic nodal line. Our work
reveals characteristic plasmon modes in a class of 2D TSMs and offers guidance for further experimental study on
NbSi$_x$Te$_2$ family materials.

\begin{figure}
\begin{centering}
\includegraphics[width=8.6cm]{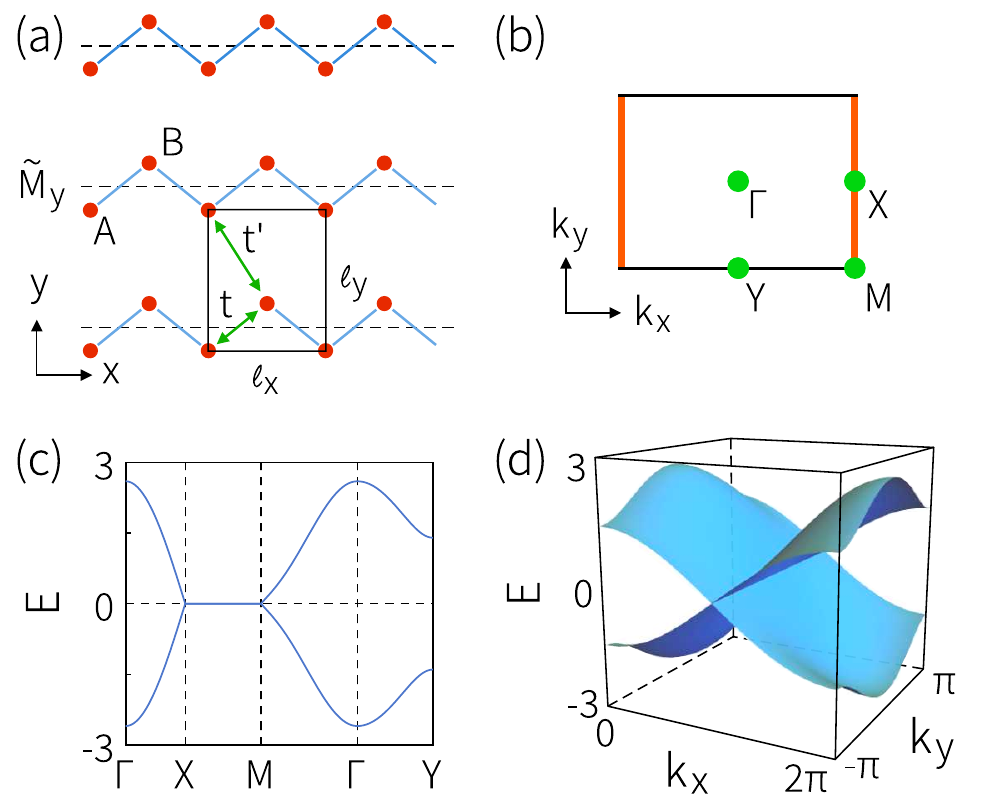}
\par\end{centering}
\caption{\label{Fig_1}(a) Schematic diagram of the lattice model. The unit cell is indicated by the black rectangle.  (b) Brillouin zone (BZ) of the model. The nodal line at the boundary of BZ is highlighted in red. (c-d) Band structure of the model. Here, energy is in unit of $t$, we set $t'=0.3t$, and $k_x$ ($k_y$) is in unit of $\ell_x^{-1}$ ($\ell_y^{-1}$).}
\end{figure}

\section{Dirac SSH model}

In order to capture the nonsymmorphic nodal line in NbSi$_x$Te$_2$ family materials, a minimal lattice model was proposed by some of the current authors in Ref.~\citep{Zhang2022_NbSiTe}. In this section, we briefly introduce this model and discuss its main features.

As illustrated in Fig.~\ref{Fig_1}(a), the model is defined on a 2D rectangular lattice, with each unit cell containing two sites $A$ and $B$. It is more illuminating by viewing the model as consisting of an array of zigzag chains running along the $x$ direction (see Fig.~\ref{Fig_1}(a)). Physically, each zigzag chain corresponds to a NbTe$_2$ chain in NbSi$_x$Te$_2$~\citep{Zhu2020_NbSiTe}. The crucial symmetries that
constrain the system are the glide mirror $\tilde{M}_y=\{M_y|\frac{1}{2}0\}$ and the time reversal symmetry $\mathcal{T}$. The glide mirror is a nonsymmorphic symmetry which involves half lattice translations along the mirror line. We shall see that these two symmetries determine the existence of the nodal line.

The model can be constructed in two steps. First, one writes down the Hamiltonian for a single zigzag chain labeled by $j$:
\begin{equation}\label{chain}
  \mathcal{H}_\text{chain}^j=\sum_{i} (t a_{i,j}^\dagger b_{i,j}+t a_{i,j}^\dagger b_{i-1,j}+h.c.),
\end{equation}
where $a_{i,j}$ and $b_{i,j}$ are the particle operators for $A$ and $B$ sites in a unit cell labeled by the index $(i,j)$. We suppress spin indices in the model, because the spin-orbit coupling strength in NbSi$_x$Te$_2$ is found to be weak~\citep{Yang2019_NbSiTe,Zhu2020_NbSiTe,Zhang2022_NbSiTe}.
This 1D model is similar to the famous SSH model~\citep{sshmodel}. However, in SSH model, the two $t$'s in the parenthesis of (\ref{chain}) are typically of different values, allowing for a dimerization pattern. In comparison, here, the two hopping amplitudes must be equal, as dictated by the  $\tilde{M}_y$ symmetry. It follows that the spectrum of model (\ref{chain}) must be gapless, with two bands crossing at a Dirac node at the BZ boundary $k=\pi/\ell_x$, where $k$ is the wave vector along the chain and $\ell_x$ is the lattice constant along $x$.

The second step is to add the interchain coupling to form a 2D model. As illustrated in Fig.~\ref{Fig_1}(a), the coupling is added between nearest sites of two neighboring chains, described by
\begin{equation}\label{inter}
  \mathcal{H}_\text{inter}=\sum_{ij}(t' a_{i,j+1}^\dagger b_{i,j}+t' a^\dagger_{i+1,j+1}b_{i,j}+h.c.).
\end{equation}
Again, the two $t'$'s in (\ref{inter}) are dictated by $\tilde{M}_y$ to be equal. In real materials, the interchain coupling is typically much smaller compared to the intrachain coupling. This will be assumed in the following discussion.

Combining the two parts gives the whole model
\begin{equation}\label{DSSH}
  \mathcal{H}=\sum_{j}\mathcal{H}^j_\text{chain}+\mathcal{H}_\text{inter}.
\end{equation}
It is important to note that adding the interchain coupling does not open a gap in the spectrum. This feature is solely dictated by symmetry. Consider the algebraic relation satisfied by $\tilde{M}_y$:
\begin{equation}
  \tilde{M}_y^2=T_{10}=e^{-ik_x \ell_x},
\end{equation}
where $T_{10}$ is the unit lattice translation along $x$ (the chain direction). Since $[\mathcal{T},\tilde{M}_y]=0$ and $\mathcal{T}^2=1$ here, we have
\begin{equation}
  (\mathcal{T}\tilde{M}_y)^2=e^{-ik_x\ell_x}.
\end{equation}
Importantly, each point on the BZ boundary path $k_x=\pi/\ell_x$ ($X$-$M$) is invariant under the \emph{anti-unitary} symmetry $\mathcal{T}\tilde{M}_y$, and $(\mathcal{T}\tilde{M}_y)^2=-1$ there. Therefore, the bands on the $X$-$M$ path must be doubly degenerate, and this dictates the presence of a nodal line along this path, as indicated in Fig.~\ref{Fig_1}(b). In the current model, this nodal line must be formed by the crossing of the two bands. This also means that the Dirac node we noticed in (\ref{chain}) is not destroyed by the interchain coupling, instead, it extends into a Dirac line at the BZ boundary in the 2D model. In order to stress this important difference from the conventional SSH model, model (\ref{DSSH}) was termed as the Dirac SSH model in Ref.~\citep{Zhang2022_NbSiTe}.

After Fourier transform to momentum space, we have
\begin{equation}\begin{split}
  \mathcal{H}(\bm k)&=t\left[
                       \begin{array}{cc}
                         0 & 1+e^{-ik_x\ell_x} \\
                         1+e^{ik_x\ell_x} & 0 \\
                       \end{array}
                     \right]\\
                     &+t'\left[
                       \begin{array}{cc}
                         0 & e^{-ik_y\ell_y}(1+e^{-ik_x\ell_x}) \\
                          e^{ik_y\ell_y}(1+e^{ik_x\ell_x}) & 0 \\
                       \end{array}
                     \right],
                     \end{split}
\end{equation}
where the first term corresponds to the intrachain coupling, and the second term corresponds to the interchain coupling. The model can be easily diagonalized and its spectrum is given by
\begin{equation}\label{Es}
  E_s(\bm k)=2s\cos\frac{k_x\ell_x}{2}\cdot \left[t^2+2tt'\cos (k_y\ell_y)+t'^2\right]^{1/2},
\end{equation}
with $s=\pm 1$. The result confirms a nodal line at $k_x=\pi/\ell_x$, where the two bands cross at zero energy. The band structure of the model is plotted in Fig.~\ref{Fig_1}(c-d), from which one can visualize the nodal line at the BZ boundary. One may expand (\ref{Es}) at a point $\bm k^*=(\pi/\ell_x,k_y)$ on the nodal line. Then, to the linear order in $\bm q$, i.e., the deviation from $\bm k^*$, we have
\begin{equation}\label{Esk}
  E_{s,\bm k^*}(\bm q)\approx-s\hbar v_F(\bm k^*) q_x,
\end{equation}
where
\begin{equation}\label{vF}
  v_F(\bm k^*)=\ell_x\left[t^2+2tt'\cos (k_y\ell_y)+t'^2\right]^{1/2}/\hbar
\end{equation}
is the Fermi velocity along $k_x$ direction at the $\bm k^*$ point. This confirms the Dirac type linear band crossing at the nodal line.

In this work, we always assume the model is around half filling, i.e., the Fermi energy is not far away from the Dirac line. One can easily calculate the density of states (DOS)
\begin{equation}
  D(E)=g\int [d\bm k]\ \delta(E-E_{s}(\bm k))
\end{equation}
for the Dirac SSH model, where $g=2$ is the spin degeneracy, and $[d\bm k]\equiv\sum_s d^2k/(2\pi)^2$ is a shorthand notation. At low energies $E$, DOS can be obtained by using Eq.~(\ref{Esk}) as
\begin{equation}\begin{split}\label{DOS}
  D(E)&= \frac{g}{\pi \Omega t}\frac{2}{\pi (1+\zeta)}K\left[\frac{4\zeta}{(1+\zeta)^2}\right]\\
  &\approx \frac{g}{\pi \Omega t}\left(1+\frac{1}{4}\zeta^2\right).
  \end{split}
\end{equation}
Here, $\Omega=\ell_x\ell_y$ is the area of a unit cell, $\zeta=t'/t<1$ is a small number by our assumption, $K$ is the complete elliptic integral of the first kind, and in the second step, we made a series expansion in $\zeta$ to the second order. As a common feature for nodal lines in 2D, the DOS is independent of energy $E$. This is similar to the usual 2D electron gas, but is distinct from 2D nodal-point TSMs (like graphene) whose DOS $D(E)\sim E$~\citep{Geim2009}. From result (\ref{DOS}), the carrier density at low doping is given by
\begin{equation}
  n(E_F)= D E_F,
\end{equation}
where $E_F$ is the Fermi energy.

Before proceeding, we comment that in 2D systems, besides nonsymmorphic symmetry, a nodal line may also be protected by a horizontal symmorphic mirror plane. For that case, the nodal line typically has the shape of a ring centered around a high-symmetry point in BZ. In comparison, here, the nonsymmorphic nodal line in Dirac SSH model is enforced to be located on a path at the BZ boundary, and this distinct shape manifests a strong in-plane anisotropy. In fact, the two different shapes of nodal lines, i.e., a ring around a point or a line traversing BZ, correspond to different topological classes. As shown in Ref.~\citep{Li2017typeii}, they are distinguished by the fundamental (homotopy) group of BZ $\pi_1(T^2)$, which describes the winding of a loop around the BZ torus. Clearly, a nodal ring does not wind around the BZ, whereas the nodal line in Fig.~\ref{Fig_1}(b) winds through the BZ once along the $k_y$ direction. This strong anisotropy of the nodal line in Dirac SSH model will manifest in the plasmon properties, as we will discuss below.

\section{Plasmons in Dirac SSH model}

Plasmons can be identified as the zeros of the dynamical dielectric function $\varepsilon\left(\boldsymbol{q},\omega\right)$ of the system. Here, we evaluate $\varepsilon\left(\boldsymbol{q},\omega\right)$ for our 2D Dirac SSH model using the random phase approximation (RPA), so that
\begin{eqnarray}\label{RPA}
\varepsilon\left(\boldsymbol{q},\omega\right) & = & 1-v_{q}\chi_{0}\left(\boldsymbol{q},\omega\right),
\end{eqnarray}
where $v_{q}=\frac{2\pi e^{2}}{\kappa q}$ the Fourier component of Coulomb potential in 2D, $\kappa$ is the background dielectric constant, and $\chi_{0}$ is the single-particle polarization function. Explicitly, we have
\begin{equation}
\begin{split}
\chi_{0}\left(\boldsymbol{q},\omega\right)  = & g \sum_{ss'}\int\frac{d^2k}{(2\pi)^2}\frac{(f_{s\boldsymbol{k}}-f_{s'\bm k+\bm q})\cdot F_{ss'}^{\boldsymbol{k},\boldsymbol{k}+\boldsymbol{q}}}{E_{s}(\bm k)-E_{s'}(\bm k+\bm q)+\hbar\omega+i\eta},
\label{chi0}
\end{split}
\end{equation}
where $f_{s\bm{k}}$ is the Fermi distribution function for the band eigenstate $|u_{s\bm k}\rangle$, $F_{ss'}^{\boldsymbol{k},\boldsymbol{k}+\boldsymbol{q}}=\left|\left\langle u_{s\boldsymbol{k}}|u_{s'\boldsymbol{k+q}}\right\rangle \right|^{2}$ is the overlap form factor, and $\eta$ is a small positive number related to the quasiparticle lifetime.

Owing to the band structure of the Dirac SSH model and the constraint from the overlap form factor, we shall see that
the contributions in $\chi_0$ from $s=s'$ and from $s\neq s'$ give distinct plasmon branches and they are separated in energy. Therefore, in the following, we shall study them separately. The former (from $s=s'$) will be referred to as intraband plasmons, and the latter (from $s\neq s'$) will be called interband plasmons.

\subsection{Intraband plasmons}

In this subsection, we examine the intraband plasmons arising from the contributions with $s=s'$.

Let's first consider the limit with $t'\rightarrow 0$, where the analysis can be greatly simplified yet the results already manifest general features of the system. In this limit, the zigzag chains are totally decoupled. Physically, in NbSi$_x$Te$_2$ family materials, this limit corresponds to the case with $x\rightarrow 1/2$.

It is important to note that for this limit, we have the form factor
\begin{equation}\label{tp0}
  F_{ss^{\prime}}^{\boldsymbol{k},\boldsymbol{k}+\boldsymbol{q}}=\delta_{ss^{\prime}},
\end{equation}
which means only intraband contribution exists, and the interband contribution is totally suppressed. Then, by substituting Eqs.~(\ref{Esk}) and (\ref{vF}) into (\ref{chi0}), one can obtain
\begin{eqnarray}
\chi_{0}\left(\boldsymbol{q},\omega\right) & = & g\frac{v_{F}}{\pi \hbar\ell_y}\frac{q_{x}^{2}}{\omega^{2}}-ig\frac{q_{x}}{\hbar\ell_y}\delta\left(\omega-v_{F}q_{x}\right),\label{eq:chi0_tp0}
\end{eqnarray}
where $v_F\equiv \ell_x t/\hbar$ is the constant Fermi velocity normal to the nodal line in this limit.

The plasmon dispersion is found by substituting (\ref{eq:chi0_tp0}) into (\ref{RPA}) and requesting $\text{Re}\ \varepsilon\left(\boldsymbol{q},\omega\right)=0$. This gives the following intraband plasmon branch
\begin{eqnarray}\label{ome}
\omega_{\boldsymbol{q}} & = & \sqrt{\frac{2g r_s}{\Omega}}v_F\sqrt{\ell_x q}\cos\theta_{\boldsymbol{q}},
\end{eqnarray}
where $r_{s}=e^{2}/\left(\kappa\hbar v_{F}\right)$ is the dimensionless fine structure constant, $q=\left|\boldsymbol{q}\right|$, and $\theta_{\boldsymbol{q}}$ is the (in-plane) polar angle for the $\bm q$ vector.

There are several notable features of this intraband plasmon branch. First, it is gapless with a $\omega\sim \sqrt{q}$ dispersion, which is characteristic for 2D systems, such as 2D electron gas and graphene. Second, the dispersion is independent of the carrier density $n$ (or the Fermi energy). This behavior is in contrast to other 2D systems, such as 2D electron gas with $\omega\sim n^{1/2}$ and graphene (nodal-point TSM) with $\omega\sim n^{1/4}$~\citep{DasSarma2009}. It is a character for a 2D nodal-line TSM. Third, as we discussed, the nonsymmorphic nodal line here exhibits a strong anisotropy. This manifests in the dispersion through the $\cos\theta_{\boldsymbol{q}}$ dependence, so that the dispersion is largest in the $x$ direction and is suppressed in the $y$ direction. Finally, due to the $r_s$ factor, the frequency explicitly involves the Planck constant $\hbar$. This is in contrast to ordinary electron liquids where the long wavelength plasma frequency is classical (does not involve $\hbar$)~\cite{Book_Electron_Liquid}. As noted in Ref.~\citep{DasSarma2009}, this feature indicates the quantum character of Dirac type fermions.

The damping and lifetime of plasmons are closely related to the single particle excitation (SPE) continuum, which can be derived from the imaginary part of the polarization function.
From the Eq.~(\ref{eq:chi0_tp0}), we see that in the limit of decoupled chains,
SPE occurs only on the line $\omega=v_{F}q_{x}$ in the $\left(\bm q,\omega\right)$ parameter space. This is in sharp contrast to other 2D systems (like 2D electron gas and graphene) where the SPE always covers an extended region~\citep{Hwang2007,Book_Electron_Liquid}. It follows that the plasmons in the Dirac SSH system can be hardly damped by SPEs and should have a long lifetime. This feature could be beneficial for possible plasmon-based applications.

Next, we add the interchain coupling $t'$ and investigate its effects on the intraband plasmons. Again, please note that we assume $\zeta=t'/t$ is small. In this case, the overlap form factor becomes
\begin{eqnarray}
F_{ss^{\prime}}^{\boldsymbol{k},\boldsymbol{k}+\boldsymbol{q}} & = & \delta_{ss^{\prime}}-\frac{1}{4}ss^{\prime}\zeta^{2}q_{y}^{2}\cos^{2}\left(k_{y}\ell_y\right),\label{eq:overlap}
\end{eqnarray}
where higher order correction terms of $O(\zeta^4)$ are ignored. One observes that the second term in (\ref{eq:overlap}) gives a correction for the intraband contribution. Meanwhile, it also makes the interband contribution nonzero, which we shall discuss in the next subsection.

\begin{figure}
\begin{centering}
\includegraphics[width=8.6cm]{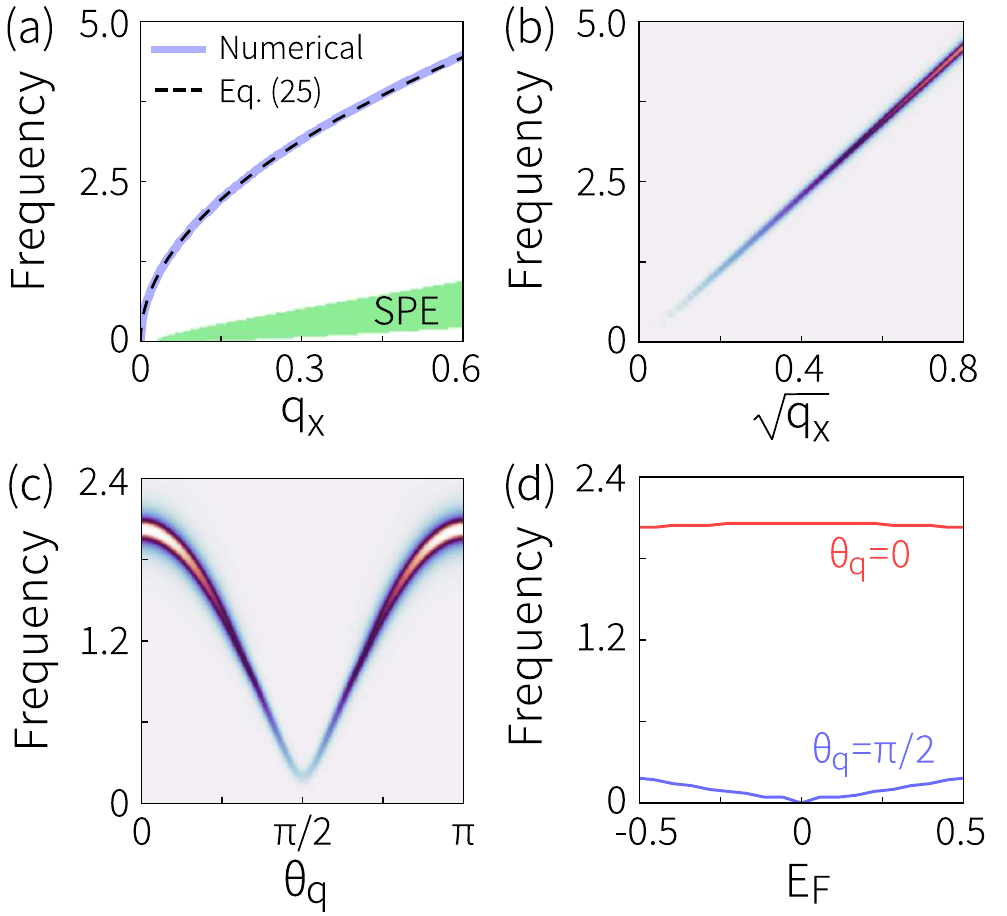}
\par\end{centering}
\caption{\label{Fig_2}Intraband plasmons of Dirac SSH model. (a) The numerical (blue solid line) and analytic (black dashed line) results of the intraband plasmon branch along $x$ direction. The SPE continuum is marked by the green colored region. (b) Plot of the energy loss function, which shows $\omega\sim\sqrt{q}$ behavior. In (a) and (b), we set $E_F=0$.  (c) Angular dependence of $\omega$ with a fixed $q=0.13/\ell_x$ and $E_F=0.5t$. (d) Plasmon frequency as a function of Fermi level $E_{F}$ with a fixed $q=0.13/\ell_{x}$.  In these figures, $\omega$, $q_x$, and $E_F$ are plotted in unit of $t/\hbar$, $\ell_x^{-1}$, and $t$, respectively. And model parameters are the same as in Fig.~1(c,d).}
\end{figure}

Substituting (\ref{eq:overlap}) into (\ref{chi0}) and considering the long wavelength limit with $q\ll k_F\equiv E_F/(\hbar v_F)$,
we find that for the intraband term
\begin{eqnarray}
\mathrm{Im}\ \chi_{0}^{\mathrm{intra}}& \approx & 2\pi g\omega \int_{\mathrm{FS}}[d\boldsymbol{k}]\,v_{s\boldsymbol{k}}^{-1}F_{ss}^{\boldsymbol{k},\boldsymbol{k}+\boldsymbol{q}}\nonumber \\
 &  & \times\delta\left[E_{s}(\bm k)-E_{s}(\boldsymbol{k}+\boldsymbol{q})+\hbar\omega\right],
\end{eqnarray}
where the integration is performed on the Fermi surface, and $v_{s\boldsymbol{k}}=\nabla_{\bm k}E_s(\bm k)/\hbar$ is the group velocity of state $|u_{s\bm k}\rangle$. From this expression, the intraband SPE continuum can be derived as
\begin{eqnarray}
\hbar\omega & = & \Lambda_{x}q_{x}+\Lambda_{y}q_{y},
\end{eqnarray}
where $\Lambda_{x}$ and $\Lambda_y$ are functions of $\bm k$, given by
\begin{eqnarray}
\Lambda_{x} & = & \ell_x t\cos\left(\frac{k_{x}\ell_x}{2}\right)\Big[1+\zeta\cos\left(k_{y}\ell_y\right)\Big],\\
\Lambda_{y} & = & -2\ell_y t^{\prime}\sin\left(\frac{k_{x}\ell_x}{2}\right)\sin\left(k_{y}\ell_y\right).
\end{eqnarray}
It follows that the SPE continuum evolves from a line in $t'=0$ limit to a fan shaped region for $t'\neq 0$ in the $(\bm q,\omega)$ space, as shown in Fig.~\ref{Fig_2}(a).

The real part of $\chi_{0}^{\mathrm{intra}}$ can be obtained as
\begin{equation}
 \mathrm{Re}\ \chi_{0}^{\mathrm{intra}}\approx g\frac{v_{F}}{\pi \hbar\ell_y}\frac{1}{\omega^{2}}\left[q_{x}^{2}-\frac{1}{4}\zeta^{2}\left(q_{x}^{2}-2k_{F}^{2}\ell_y^2 q_{y}^{2}\right)\right].
\end{equation}
Then, the dispersion of long-wavelength intraband plasmons in the presence of interchain coupling is
\begin{eqnarray}
\omega_{\boldsymbol{q}} & = & \sqrt{\frac{2gr_s}{\Omega}}v_F\sqrt{\ell_x q}\left[\left(1-\frac{1}{4}\zeta^{2}\right)\cos^{2}\theta_{\boldsymbol{q}}\right.\nonumber \\
 &  & \qquad\qquad\qquad \left.+\frac{1}{2}\zeta^{2}k_{F}^{2}\ell_y^2\sin^{2}\theta_{\boldsymbol{q}}\right]^{1/2}.\label{eq:plasmon2}
\end{eqnarray}
One can see that the interchain coupling (nonzero $\zeta$) gives a correction to the dispersion. In the limit of $\zeta\rightarrow 0$, this expression reduces to Eq.~(\ref{ome}), as expected. The previously noted features, including the
$\sqrt{q}$ dependence, the strong anisotropy, and the quantum character (involving $\hbar$ explicitly) are maintained after including the interchain coupling. The strongest dispersion of these plasmons is still along the $x$ direction, i.e., the direction normal to the nodal line. Along this direction, $\theta_{\boldsymbol{q}}=0$ and we have
\begin{eqnarray}
\omega_{\boldsymbol{q}} & = & \sqrt{\frac{2gr_s}{\Omega}\left(1-\frac{1}{4}\zeta^{2}\right)}v_F\sqrt{\ell_x q},
\end{eqnarray}
which is still independent of the carrier density $n$. The interchain coupling only gives a small correction to the frequency. Meanwhile, for plasmons propagating along the direction of nodal line, i.e, $\theta_{\boldsymbol{q}}=\pi/2$, we have
\begin{equation}
  \omega_{\boldsymbol{q}} = \sqrt{gr_s}\zeta v_F k_F \sqrt{\ell_y q}.
\end{equation}
Therefore, with the interchain coupling, the intraband plasmons in this direction acquire finite frequency.
Moreover, due to the $k_F$ factor, the plasmon frequency scales linearly with $n$ (or $E_F$). In other words, at finite interchain coupling, there is a crossover in the scaling of intraband plasmon frequency from $\omega\sim n^0$ to $\omega\sim n$, when $\theta_{\bm q}$ varies from $0$ to $\pi/2$. Such a crossover behavior clearly reflects the strong anisotropy of the nonsymmorphic nodal line.

These features are verified by our numerical calculations on the Dirac SSH model. In the calculation, the plasmon modes are extract from the peaks of the energy loss function
\begin{eqnarray}
L\left(\boldsymbol{q},\omega\right) & = & \mathrm{Im}\left[-\frac{1}{\varepsilon\left(\boldsymbol{q},\omega\right)}\right].\label{eq:loss}
\end{eqnarray}
As shown in Fig.~\ref{Fig_2}(a), the result from the numerical calculation agrees very well with the analytic formula (\ref{eq:plasmon2}). The plasmons are quite separated from the SPE continuum which only spans a small region. In Fig.~\ref{Fig_2}(b), we plot the frequency versus $q^{1/2}$, so that one can clearly see the $\omega\sim q^{1/2}$ dispersion. Figure \ref{Fig_2}(c) shows the angular dependence of plasma frequency at a fix wave vector magnitude. The frequency is maximum at direction normal to the nodal line and is minimum along the nodal line, reflecting the strong anisotropy of the system. Finally, in Fig.~\ref{Fig_2}(d), we plot $\omega$ as a function of $E_F$ for different propagation directions, which confirms the scaling behavior discussed above.

\begin{figure}
\begin{centering}
\includegraphics[width=8.6cm]{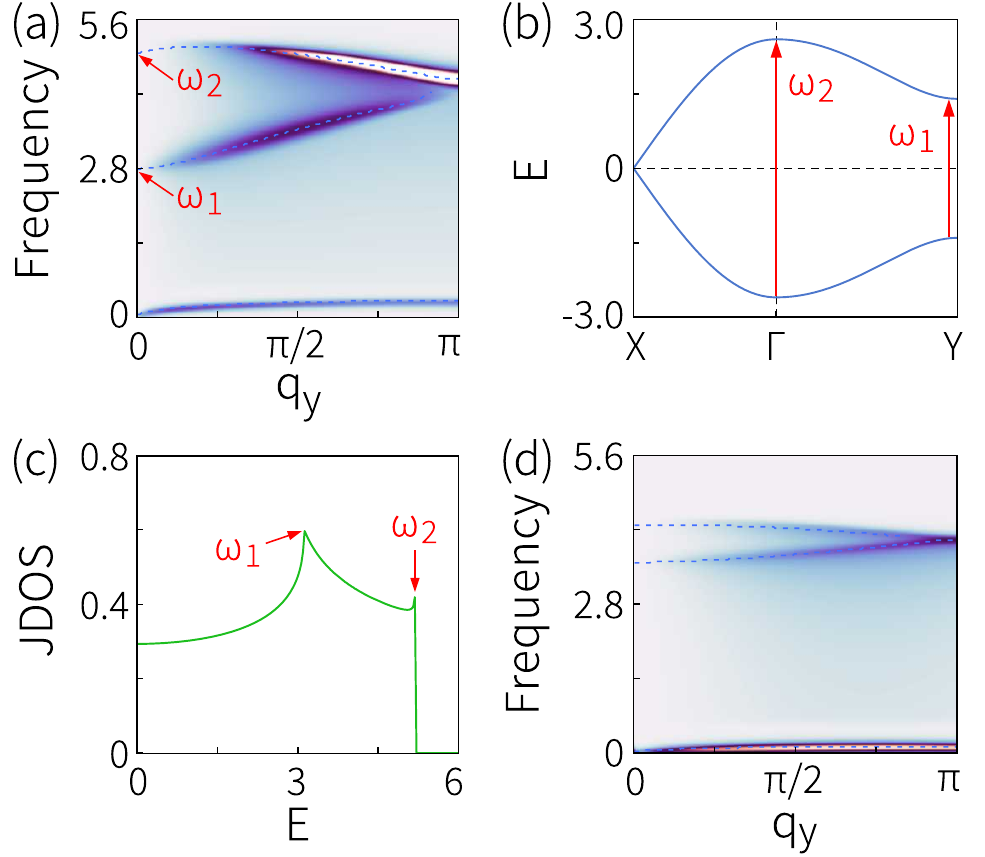}
\par\end{centering}
\caption{\label{Fig_3} (a) Plasmon spectrum along the nodal line. The two upper branches are the interband plasmons. (b)  The two frequencies $\omega_1$ and $\omega_2$ associated with the van Hove singularities at $Y$ and $\Gamma$ are marked by the red arrows. (c) Joint density of states (JDOS) of the model. The two frequencies manifest as singularities in the plot. In (a-c), the model parameters are the same as in Fig.~1(c,d) (with $\zeta=0.3$). $q_y$ and JDOS are plotted in units of $\ell_y^{-1}$ and $1/(\Omega t)$. (d) is the same as panel (a) but with a smaller interchain coupling $\zeta=0.1$. }
\end{figure}

\subsection{Interband plasmons}

From the overlap form factor in (\ref{eq:overlap}), we notice that along the $x$ direction, only intraband plasmons exist. However, for other directions, especially the $y$ direction, interband contributions with $s\neq s'$ can exist when there is nonzero interchain coupling $t'$. For these interband modes, $\chi_0$ may involve virtual transitions between states away from the nodal line, so it is difficult to get simple analytic result. In the following, we shall proceed with numerical calculation of the Dirac SSH model.

Figure \ref{Fig_3}(a) shows the obtained energy loss function for $\bm q$ along the nodal-line ($y$) direction, where such interband plasmons are most pronounced. One observes that besides the gapless intraband branch at low frequencies, there are two other gapped branches at much higher frequencies. They correspond to the interband plasmons.

As we mentioned, these interband modes are related to interband virtual transitions in the band structure. Actually, one can see that the long wavelength limit of the two branches encode the information of the van Hove singularities of the Dirac SSH band structure. As indicated in Fig.~\ref{Fig_3}(b), in the $q\rightarrow 0$ limit, the virtual transitions are vertical, and there are two dominant frequencies $\omega_1$ and $\omega_2$ corresponding to the van Hove singularities at  $Y$ and $\Gamma$ of BZ. It is more convenient to visualize these two frequencies as corresponding to the singularities in the joint density of states (JDOS), as shown in Fig~\ref{Fig_3}(c). In the plasmon spectrum, $\omega_1$ and $\omega_2$ are exactly the limiting values of the two interband branches in the $q\rightarrow 0$ limit. (Here, the intensity of the peak vanishes, because the form factor is zero in this limit.) With increasing $q$, the frequency of the upper branch decreases and the lower branch increases, and the two merge at $q=\pi/\ell_y$, corresponding to the interband transitions from $\Gamma$ to $Y$ or from $Y$ to $\Gamma$, owing to the particle-hole symmetry of the model.

Another feature to be noted is that the band width of these two interband branches is on the order of interchain coupling $t'$. This can be easily understood by noting that this width scales with the electronic band energy variation along $k_y$ which is governed by $t'$. In Fig.~\ref{Fig_3}(d), we plot the plasmon spectrum at a decreased value of $t'$ as compared to Fig.~\ref{Fig_3}(a). Indeed, one observes that the interband branches become flatter.

\section{Material result}

\begin{figure}
\begin{centering}
\includegraphics[width=8.6cm]{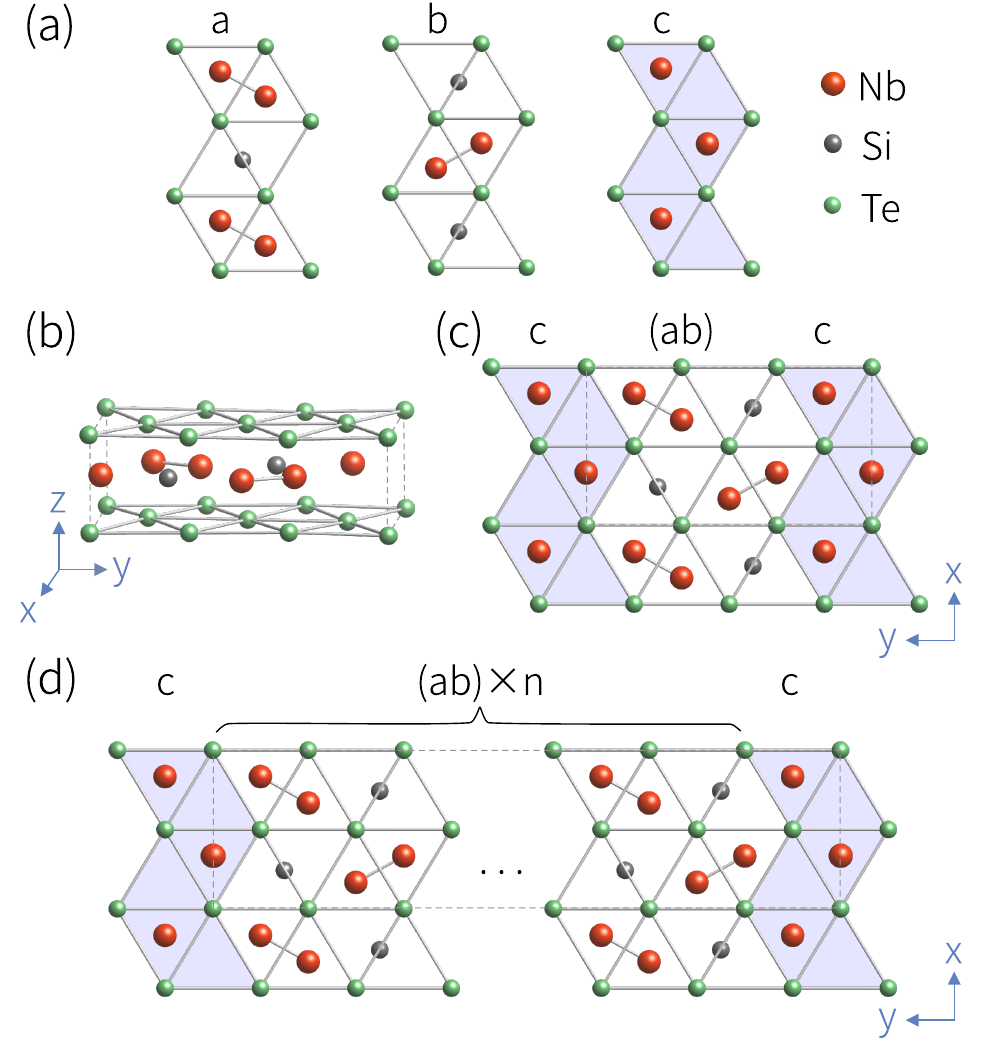}
\par\end{centering}
\caption{\label{Fig_4}(a) Building blocks (a, b, and c chains) of $\mathrm{Nb}\mathrm{Si}_{x}\mathrm{Te}_{2}$ family materials. See the discussion in the text. (b-c) Side and top views of Nb$_3$SiTe$_6$, corresponding to the $n=1$ case. The dashed lines mark the unit cell. (d) Structure of general configuration (ab)$_n$c, $n\geq 1$, where $n$ copies of (ab) chains are inserted between two c chains. }
\end{figure}

We have studied the plasmonic properties of 2D Dirac SSH model. In this section, we shall see whether these properties can manifest in the real material system, i.e., the monolayer $\mathrm{Nb}\mathrm{Si}_{x}\mathrm{Te}_{2}$ family materials.

Figures \ref{Fig_4}(b-c) show the crystal structure of Nb$_3$SiTe$_6$ (i.e., $x=1/3$). The structure consists of three atomic layers. The Nb and Si atoms form the middle layer, which are sandwiched by two layers of Te atoms. Within the 2D plane, the structure can be viewed as constructed by three basic building blocks, which are conventionally named as a, b, and c chains, as illustrated in Fig.~\ref{Fig_4}(a). a and b chains contain Si atoms, and they share the same composition of NbSi$_{1/2}$Te$_2$. In comparison, the c chain does not contain Si, so it has a composition of NbTe$_2$. a and b chains are connected by the glide mirror symmetry $\tilde{M}_y$, and they always stick together. In terms of these building blocks,  Nb$_3$SiTe$_6$ in Fig.~\ref{Fig_4}(c) has the configuration of (ab)$_1$c. Other members of the $\mathrm{Nb}\mathrm{Si}_{x}\mathrm{Te}_{2}$ family are obtained by repeating more (ab) chain units between the c chains, so they have the general configuration of (ab)$_n$c~\citep{Zhu2020_NbSiTe}, as shown in Fig.~\ref{Fig_4}(d).
In this picture, their general chemical formula may also be written as (Nb$_2$SiTe$_4$)$_n$(NbTe$_2$).

\begin{figure}
\begin{centering}
\includegraphics[width=8.6cm]{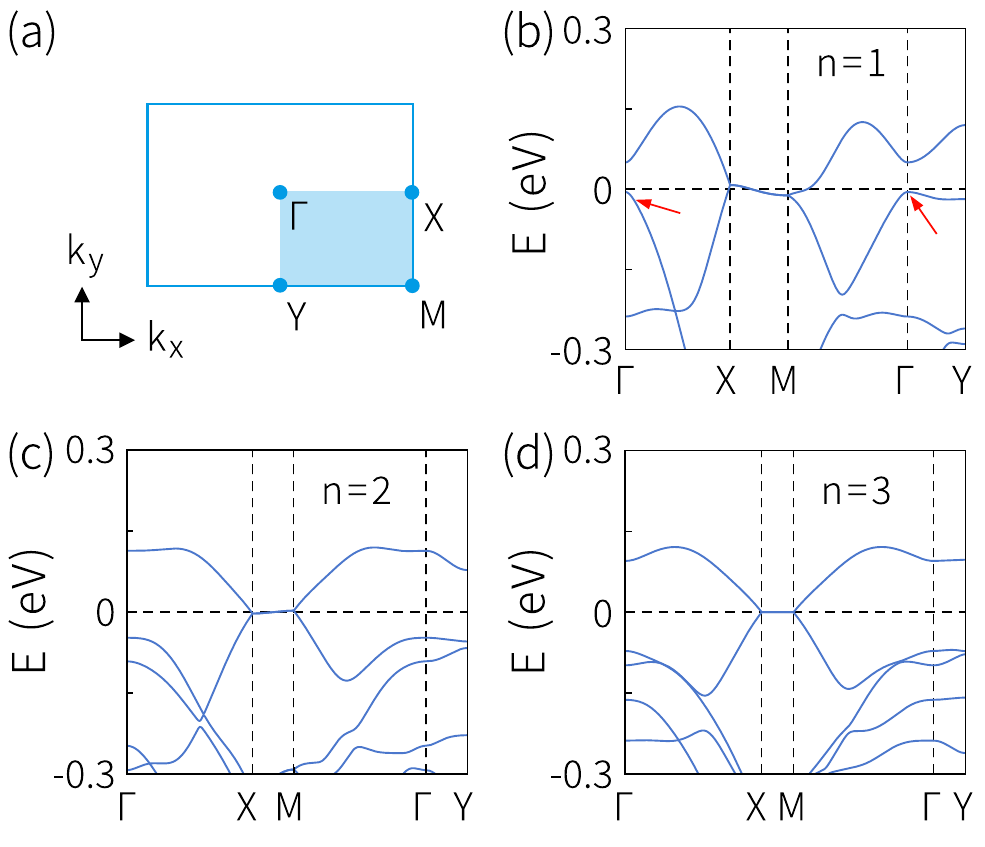}
\par\end{centering}
\caption{\label{Fig_5}(a) BZ of $\mathrm{Nb}\mathrm{Si}_{x}\mathrm{Te}_{2}$ family materials. (b-d) Band structures of (Nb$_2$SiTe$_4$)$_n$(NbTe$_2$) with $n=1,2,3$.}
\end{figure}

We perform first-principles calculations on the three member materials with $n=1,2,3$. The calculation method is given in Appendix. Previous works showed that the spin-orbit coupling is weak in these materials~\citep{Yang2019_NbSiTe,Zhu2020_NbSiTe,Zhang2022_NbSiTe}, so it is neglected in our calculation. The obtained band structures are plotted in Fig.~\ref{Fig_5}, which agree with previous results~\citep{Zhu2020_NbSiTe}. One observes that the three band structures share similar features, with a nonsymmorphic nodal line close to the Fermi level on the $X$-$M$ path, formed by the crossing of two bands. Previous studies showed that the low-energy states are mainly from the c chains~\citep{Zhu2020_NbSiTe,Wang2021_NbSiTe}. The intrachain coupling is strong, making sizable dispersion along the $x$ direction (along the chain). The interchain coupling is relatively weak and it naturally decreases with increasing $n$.

\begin{figure}[t]
\begin{centering}
\includegraphics[width=8.6cm]{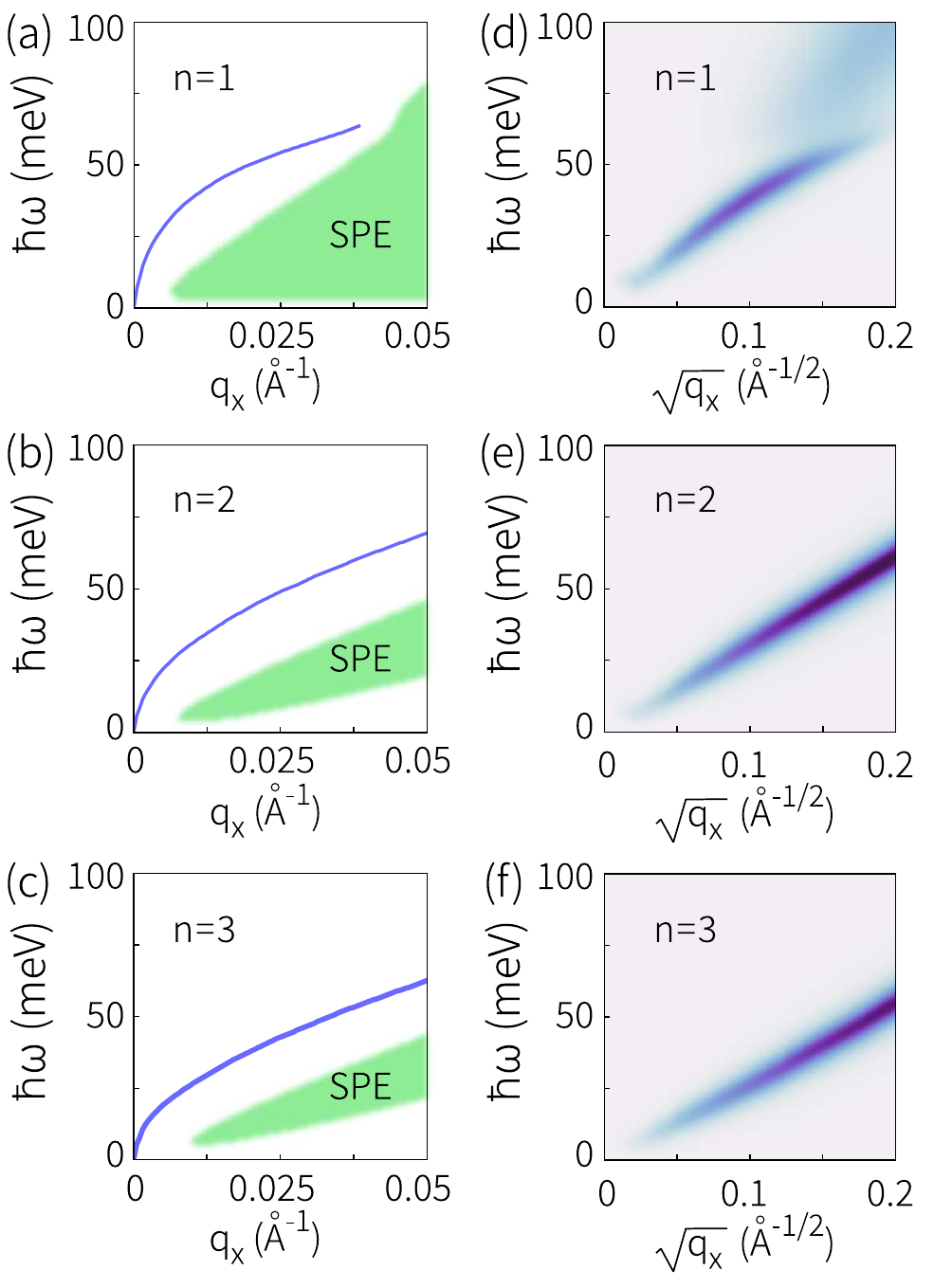}
\par\end{centering}
\caption{\label{Fig_6} (a-c) Plasmon spectra along $x$ (normal to the nodal line) for $\mathrm{Nb}\mathrm{Si}_{x}\mathrm{Te}_{2}$ family materials. The SPE regions are marked by the green color. (d-f) show the plot of energy loss function in this direction and the plot is versus $\sqrt{q}$ to explicitly show the scaling. We take $E_F=0$ (i.e., undoped case) in these calculations. }
\end{figure}

\begin{figure*}
\begin{centering}
\includegraphics[width=17.6cm]{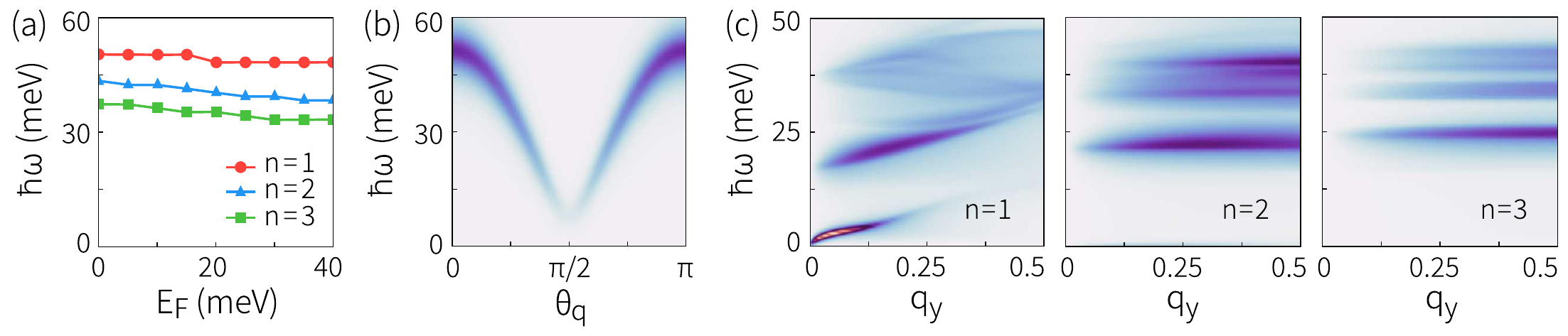}
\par\end{centering}
\caption{\label{Fig_7}(a) Frequency of intraband plasmon in $\mathrm{Nb}\mathrm{Si}_{x}\mathrm{Te}_{2}$ as a function of Fermi level $E_F$. Here, $q=0.02~\rm\AA^{-1}$ and $\theta_{\boldsymbol{q}}=0$. (b) Angular dependence of the intraband plasmons with a fixed $q=0.02~\rm\AA^{-1}$. Here, we set $E_F=-50$~meV. (c) Plasmon spectra along $y$ direction for $n=1,2,3$ cases. We set $E_F=0$ and the $q_y$ is in unit of $2\pi\ell_y^{-1}$. }
\end{figure*}

Based on the first-principles band structures, we compute their plasmon spectra in RPA. First, consider the plasmon dispersion along $x$, i.e., the direction normal to the nodal line. The results are plotted (versus $q^{1/2}$) in Fig.~\ref{Fig_6}. One can see that there is a single gapless branch with $\omega\sim q^{1/2}$ in this direction, representing the intraband plasmons, consistent with our model result in Fig.~\ref{Fig_2}. The plasmon peaks in Fig.~\ref{Fig_6}(e,f) are quite sharp, indicating their weak Landau damping and long lifetime. Indeed, our calculation shows that the SPE continuum for $n=2$ or $3$ is far separated from the plasmon branch. In comparison, the plasmon peaks for $n=1$ are smeared out for $\hbar\omega$ above 50 meV [Fig.~\ref{Fig_6}(d)]. This is due to the overlap with the SPE region [see Fig.~\ref{Fig_6}(a)]. The SPE continuum for $n=1$ spans a much larger region, because the material's band structure has a stronger deviation from the Dirac SSH model, as indicated by the red arrows in Fig.~\ref{Fig_5}(b).

In Fig.~\ref{Fig_7}(a), we show the variation of the intraband plasmon frequency with Fermi energy. One observes that at a fixed wave vector $q_x$, the frequencies are almost independent of $E_F$, which confirms the scaling behavior we find in the model study.

In Fig.~\ref{Fig_7}(b), we plot the angular dependence of intraband plasmons, showing a behavior consistent with the model result in Fig.~\ref{Fig_2}(c).  It confirms that the dispersion is strongest along $x$ (i.e., normal to the nodal line) and is suppressed along $y$ (parallel to the nodal line).

Next, we plot the plasmon spectra along the $y$ direction in Fig.~\ref{Fig_7}(c). One observes that besides the intraband plasmon branch at low frequency (which are almost entirely suppressed for $n=2,3$, indicating their very small interchain coupling), there are
interband branches at higher frequencies. This agrees with the model result. Nevertheless, there appear to be more than two interband branches, especially for $n=2$ and $3$. This is due to the presence of other nearby bands in Fig.~\ref{Fig_5}, which also contribute to the interband transitions. Consistent with our model analysis, one can see that the width of the interband branches decreases with $n$. For $n=2$ and $3$, these branches become very flat, again reflecting the weak interchain coupling.

\begin{figure}[b]
\begin{centering}
\includegraphics[width=8.6cm]{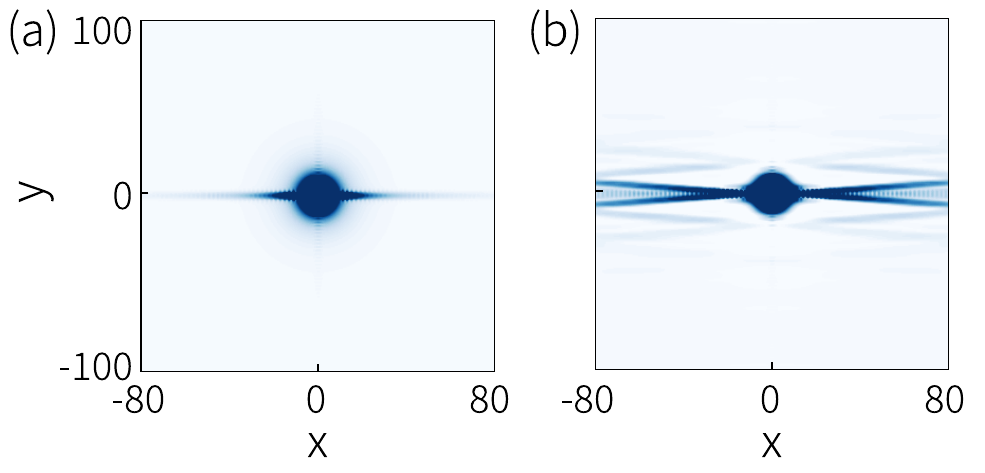}
\par\end{centering}
\caption{\label{Fig_8} Screening charge density $\delta n(\boldsymbol{r})$ induced by a charge impurity at the origin. (a) $\zeta=0$ and (b) $\zeta=0.3$. Here, $x$ and $y$ are plotted in unit of $\ell_x$ and $\ell_y$. The model parameters are the same as in Fig.~1(c,d) and we take $E_F=50$~meV.}
\end{figure}

\section{Discussion and Conclusion}

We have discussed the distinct features of plasmons in the class of nonsymmorphic nodal-line TSMs. The example materials used in our study, namely the NbSi$_x$Te$_2$ family materials, already exist. Their high-quality 2D ultrathin layers have been demonstrated in experiment~\citep{Hu2015}. In addition, there exist several other isostructural materials, including TaSi$_x$Te$_2$, NbGe$_x$Te$_2$, and TaGe$_x$Te$_2$, with the similar nodal-line band structure~\citep{Li2018_NbSiTe,evain1994modulated,van1994superspace}. We expect our theory can be readily tested in these materials.

As for probing the plasmon excitations, there are several established experimental techniques, such as electron energy loss spectroscopy (EELS)~\citep{Politano2015,Xun2017}, inelastic helium atom scattering, and transmission electron microscopy. Particularly, the high-resolution EELS provides a powerful tool for mapping out the plasmon spectrum. It was demonstrated that the technique can reach a resolution of a few meV and the detected frequency range can reach  $\sim 1$~eV~\citep{Xun2017,Xue2021}. Hence, our predicted features in 2D NbSi$_x$Te$_2$ should be easily detectable in experiment.

In addition, we mention that besides plasmons, the unusual dielectric function for nonsymmorphic nodal-line TSMs may also manifest in the screening charge distribution induced by a charge impurity. Consider a charge impurity $n^\text{ext}$ in a 2D system. Its induced screening charge density $\delta n$ can be expressed as
\begin{equation}
  \delta n(\bm r)=\int \frac{d\bm q}{(2\pi)^2}\Big[\frac{1}{\varepsilon(\bm q,0)}-1\Big]n^\text{ext}(\bm q)e^{i\bm q\cdot \bm r}.
\end{equation}
For usual 2D electron gas, the induced charge density $\delta n$ exhibits the well-known Friedel oscillation, with $\delta n(\bm r)\sim \cos(2k_F r)/r^2$. The oscillation wave vector $2k_F$ arises from the singularity in the static dielectric function $\varepsilon(\bm q,0)$. Now, consider the Dirac SSH model. For the $t'\rightarrow 0$ limit, the Fermi surface (at $E_F\neq 0$) are given by two straight lines parallel to the nodal line. From Eq.~(\ref{tp0}), singularities in $\varepsilon(\bm q,0)$ arise only from the nesting vector restricted to each piece of the Fermi surface. Importantly, the special Fermi surface here gives not one but a continuum of nesting wave vectors $q\hat{y}$ along the $y$ direction. After Fourier transform to real space, the induced charge density should exhibit peaks on a straight line along $x$ through the charge impurity, as shown in Fig.~\ref{Fig_8}(a). Increasing $t'$ to a nonzero but small value, we find that the distribution of singular wave vectors evolve from a single line to two lines forming a cross at $\bm q=0$. Accordingly, the screening charge density peaks also transform from the horizontal line in Fig.~\ref{Fig_8}(a) to two almost straight lines through the charge impurity as in Fig.~\ref{Fig_8}(b).

In conclusion, we discover interesting plasmon properties in a class of 2D nonsymmorphic nodal-line TSMs. Using the Dirac SSH model, we show that the system possess two kinds of plasmons: the gapless intraband plasmons and the gapped interband plasmons. The intraband plasmons has a $q^{1/2}$ dependence. The dispersion and the carrier density scaling are highly anisotropic. Normal to the nodal line, the dispersion is strongest and is independent of carrier density; whereas along the nodal line, the dispersion is largely suppressed and is linear in the carrier density. The interband plasmons have a quite flat dispersion, with band width scaling with the interchain coupling. Their long wavelength limits are connected with transitions between van Hove singularities of the band structure. Most these plasmons are separated from the SPE continuum, so they should have a weak Landau damping. The revealed features are also verified in the NbSi$_x$Te$_2$ family materials via first-principles calculations. Our work reveals new physical consequences of a topological state of matter. The predicted physics will also motivate further studies on 2D nonsymmorphic nodal-line TSMs and on NbSi$_x$Te$_2$ family materials.

\begin{acknowledgments}
The authors thank D. L. Deng for helpful discussions. This work is supported by Singapore MOE AcRF Tier 2 (MOE2019-T2-1-001), the National Key R\&D Program of China (Grant No.~2020YFA0308800), and the NSF of China (Grants Nos.~12234003, 12061131002, 11547200).
\end{acknowledgments}

\appendix
% \counterwithin{figure}{section}
% \counterwithin{table}{section}
% \counterwithin{equation}{section}
% begin{appendix}
   \renewcommand{\theequation}{A\arabic{equation}}
   \setcounter{equation}{0}
   \renewcommand{\thefigure}{A\arabic{figure}}
   \setcounter{figure}{0}
   \renewcommand{\thetable}{A\arabic{table}}
   \setcounter{table}{0}

\section{Calculation method}
The electronic structures of the $\mathrm{Nb}\mathrm{Si}_{x}\mathrm{Te}_{2}$ were calculated based on the density functional theory (DFT) using the Vienna ab-initio Simulation Package~\citep{VASP1,VASP2,VASP3}. The ionic potentials were treated with the projector augmented way pseudopotentials~\citep{PAW}, with $4p^{6}5s^{1}4d^{4}$, $3s^{2}3p^{2}$, and $5s^{2}5p^{4}$ valence electron configurations for Nb, Si, and Te atoms, respectively. The exchange-correlation energy was treated with the generalized gradient approximation~\citep{GGA} in the scheme of Perdew-Burke-Ernzerhof (PBE)~\citep{PBE} approach. The plane wave cutoff energy was set to be 400 eV and the convergence thresholds for energy and force were chosen to be $10^{-7}$~eV and $0.001\,\mathrm{eV/\mathring{A}}$, respectively. $\Gamma$-centered $k$-points meshes with size $10\times6\times1$ for (ab)$_{1}$c and $10\times4\times1$ for (ab)$_2$c and (ab)$_3$c were used for BZ sampling. After obtaining the band structure, ab-initio tight-binding models were constructed by using the WANNIER90 package~\citep{WANNIER90_CPC}. The $p$ orbitals of Te atoms and $d$ orbitals of Nb atoms were used as initial input of local basis. The Wannier model were used to calculate the dynamical dielectric function in RPA.

\bibliographystyle{apsrev4-2}
\bibliography{ref}

\end{document}